\documentclass[aps,preprint]{revtex4-1}  
\usepackage{amssymb}
\usepackage{amsfonts}
\usepackage{amsmath}
\usepackage{array}
\usepackage{mathrsfs}
\usepackage{graphicx}
\usepackage{flushend}
\usepackage{float}
\usepackage{subfigure}
\usepackage{multirow}
\usepackage{booktabs}
\usepackage{bm}
\usepackage{hyperref}
\hypersetup{
	colorlinks=true,
	linkcolor=blue,
	urlcolor=blue,
	citecolor=red,
	linktoc=all
}

\begin{document}

\title{Macroscopicity of quantum superposition}

\author{Xiao-Fu Peng}
\email{E-mail: pengxf@hust.edu.cn}
	
\affiliation{Hubei Key Laboratory of Gravitational and Quantum Physics, School of Physics, Huazhong University of Science and Technology, Wuhan 430074, China}

\date{\today}
	
\begin{abstract}
	In this paper, we define a universal measure of macroscopicity $\beta$ in the form of, ({\em experimentally observed coherence time})/({\em characteristic time required to repeatably distinguish the components of a quantum superposition state}), to divide the quantum/classical (Q-C) boundary. The analysis shows that, for a matter-wave interferometer, covering systems from electron to macromolecular, the measure of its macroscopicity is expressed by $\beta=p\theta d/8\hbar$, and it consistently yields $\beta<1$ by investigating the coherence data reported in the literature. The result implies that limited advancement has been made toward exploring the Q-C boundary in such experiments. Additionally, for a Ramsey-like interferometer, the measure of its macroscopicity is expressed as $\beta=8\pi^3\alpha\nu_0r_s^2\nu_s^2T/ (9c^2)$. After analyzing the reported experimental data, we find that a maximum value of $\beta \sim 300$ is achieved in the vertical atomic fountain experiments. Therefore, searching the Q-C boundary in the atomic fountain may be a more proper direction. 
\end{abstract}

    \maketitle

\section{Introduction}

In the century anniversary of quantum mechanics, it is commemorative to discuss a fundamental question about the quantum/classical (Q-C) boundary, since the celebrated Schr{\" o}dinger's cat still puzzles us \cite{1935Schrodinger}. Quantum superposition behavior, as known in the microscopic domain, has been widely verified by experiments, while in the macroscopic realm, our daily experience tells us a cat cannot be alive and dead at the same time. So a natural question is where the boundary between them is. Due to the success of quantum mechanics, one may ask to what extent the macroscopic superposition can hold.

It seems that the decoherence theory can respond to these questions. The phenomenon on a detector only captures the information at a coarse level, while the fine and huge one behind the pointer has been inevitably traced by the environment \cite{2003Zurek,2019Schlosshauer}. However, the complexity of the decoherence process usually prevents us from acquiring an exact reason that prompts the quantum-to-classical transition in practice. Besides, it is also hard to believe that the theory within standard quantum mechanics can completely solve this problem \cite{2003Adler}. Based on these, people began to select a quantity, which can empirically distinguish the characteristics of both quantum states and classical ensembles, as an indicator for demarcating the Q-C boundary artificially \cite{2018macroscopic}. Such quantities can be employed to simply measure the macroscopicity of a system. Maybe it can also provide some clues toward a unified description in both quantum and classical worlds.

A. J. Leggett first proposed a macroscopic indicator, which is defined as disconnectivity $D$, through the entropy relation of subsystems \cite{1980Leggett}. He then showed that the phenomena in superfluids had a small $D$, while large $D$ appeared in the entanglement system of multiple particles. It conforms to our intuition that the classical result is displayed only when a quantum particle entangles with the macroscopic detector. After that, people attempted to connect the macroscopicity with a large entangled state in a variety of ways \cite{2002Dur,2012Frowis}. Other approaches that can produce a reasonable macroscopic indicator include those through the scale and mass of a quantum system \cite{2008Cavalcanti,2013macro,2020macro}, or directly from the interference utility \cite{2004Bjork} and classical information in measurements \cite{2014_Sekatski,2018_Sekatski}. However, most of the existing indicators have different scopes in application, so they capture only a limited set of characteristics within specific types of systems. For example, the mass-based indicator is not well-suitable for judging the superposition in spin esembles.

In this paper, we introduce a universal measure of macroscopicity, $\beta$, defined as the ratio between the experimentally observed coherence time $T$ and the characteristic time $\tau$ required to repeatably distinguish the components of the superposition state: 
\begin{equation}\label{def-beta}
    \beta=\frac{\text{ Experimentally  observed  coherence  time}\ T}{\text{ Repeatably distinguishing time}\ \tau}.
\end{equation}
This measure originates from a new model in Ref. \cite{R2model}, which is proposed to address the quantum measurement problem. For the components of any quantum superposition, there is a possible repeatably distinguishing time $\tau$ in principle. Consequently, the measure $\beta$ could be used to discuss the macroscopicity of such quantum systems.

Here we will show specifically how to employ the measure in two kind of quantum systems, i.e., $\beta=p\theta d/8\hbar $ for matter-wave interferometer and $\beta=8\pi^3\alpha\nu_0r_s^2\nu_s^2T/ (9c^2)$ for Ramsey-like interferometer.  We find that the macroscopicity measure in macromolecular interferometers is smaller than 1, i.e. $\beta<1$, while in atomic fountains one has $\beta\sim 370$. The result shows that searching the Q-C boundary in the atomic fountain may be a more proper direction, which contrasts with our intuition. 

This paper is organized as follows: In Sec. \ref{beta}, we introduce the universal macroscopicity measure, and show the repeatably distinguishing time $\tau$ for the matter-wave and the Ramsey-like interferometers. In Sec. \ref{interferometer} and Sec. \ref{atomic fountain}, we employ the measure $\beta$ to analyze the macroscopicity of matter-wave and Ramsey-like interferometers, through the experiment data, respectively. We conclude the result and give a prospect in Sec. \ref{conclusion}. 

\section{The Universal Macroscopicity Measure}\label{beta}

Obviously, the repeatably distinguishing time $\tau$ plays a central role in our core formula Eq. \eqref{def-beta}. To better understand the physical meaning of this quantity, let us consider a macroscopic example: a cat that can occupy either of two distinct positions \(|\text{here}\rangle\) or \(|\text{there}\rangle\). In the classical world, one can determine whether the cat is \(|\text{here}\rangle\) or \(|\text{there}\rangle\) without significantly affecting its state. This means the two macroscopic position states are repeatably distinguishable.
By analogy, we can generalize this idea to the quantum domain. Given two quantum states, such as energy eigenstates \(|E_1\rangle\) and \(|E_2\rangle\), we define them as repeatably distinguishable if there exists a method to determine whether the system is \(|E_1\rangle\) or \(|E_2\rangle\) without significantly disturbing the state. The repeatably distinguishing time \(\tau\) is then naturally defined as the minimal time required to carry out such a process. 

Finally, by comparing the repeatably distinguishing time $\tau$ with the experimentally observed coherence time $T$, we arrive at the universal macroscopicity measure $\beta$, as shown in Eq. \eqref{def-beta}. Before proceeding to detailed applications, let us qualitatively examine some fundamental implications of $\beta$. 

Consider a measurement process in which a quantum system is entangled with the position of a classical pointer. In this case, we have $\tau\rightarrow 0$, since one can repeatedly distinguish where the pointer is without any effort. As a result, even just after a very short measurement time $T$, the macroscopicity diverges $\beta\rightarrow\infty$, which indicates the emergence of classical behavior. 
In contrast, consider a stationary quantum state, such as the ground state of a valence electron in an atom. In this scenario, it is fundamentally impossible to determine the electron’s position without applying a strong external probe. However, any such attempt would inevitably destroy the state. Hence, we have $\tau\rightarrow\infty$, and the macroscopicity parameter vanishes $\beta\rightarrow 0$, no matter how long $T$ the stationary state goes through.

So far, we have examined extremely large and small limits of $\beta$, which conforms to our intuitive impression of quantum and classical systems. The following examples demonstrate two kinds of quantum systems where $\beta$ is finite and not zero, so we can utilize them to quantize the macroscopicity in practical experiments. 

Let us firstly consider the matter-wave interferometer, and the basic setup is as follows: A particle of mass $M$ and momentum $p=Mv$ to fly freely from an origin $O$ toward a plate with two parallel slits, located symmetrically with respect to $O$.  The distance of $O$ to the plate is $L$. The two slits are separated by a distance $D$, and each slit has a width $d$ (see Fig. \ref{slit} below). The observed coherence time should be the flying time of particles,
\begin{equation}\label{mw-T}
    T=\frac{L}{v}.
\end{equation}

According to the derivation in Ref. \cite{R2model}, the characteristic time required to repeatably distinguish two momentum trajectories is given by
\begin{equation}\label{mw-tau}
    \tau=\frac{8\hslash L}{vp\theta d},
\end{equation}
where $\theta$ denotes the angle between the two paths. A few additional remarks can be made about the physical meaning of this time scale. Actually, $p\theta$ represents the momentum difference of the particles on two paths, and $vd/L$ can be regarded as the momentum uncertainty of the particle passing-through the slits in some sense. Larger values of either quantity enhance the repeatable distinguishability of the two paths, thereby reducing $\tau$.

Another practical example is the Ramsey interferometer, where the typical components of a superposition are the hyperfine states within the ground state of an atom. The characteristic time $\tau$ required to repeatably distinguish these components can also be derived (for a simple derivation, see Appendix \ref{rdt}), and is given by
\begin{equation}\label{Ram-tau}
	\tau=\frac{9}{8\pi^3\alpha\nu_0}\left(\frac{c}{r_s\nu_s}\right)^2,
\end{equation}
where $\alpha$ is the fine structure constant, $\nu_0$ is the energy gap and, $\nu_s$ and $r_s$ denote the system frequency and size of the host atom, respectively

The physical interpretation of the distinguishing time $\tau$ presented above is clear: the larger the energy gap $\nu_0$, the easier it is to distinguish between two states; the larger the system scales, represented by $\nu_s$ and $r_s$, the more resilient the system becomes against disturbation. Therefore, the final $\tau$ should be inversely proportional to some powers of these parameters, as we desired. In typically classical systems, these parameters increase dramatically, leading to the emergence of the macroscopic world.

Now, let us measure the macroscopicity of existing experiments.

\section{Macroscopicity of Matter-wave Interferometer}\label{interferometer}
\begin{figure}
	\centering
        \includegraphics[width=0.5\linewidth]{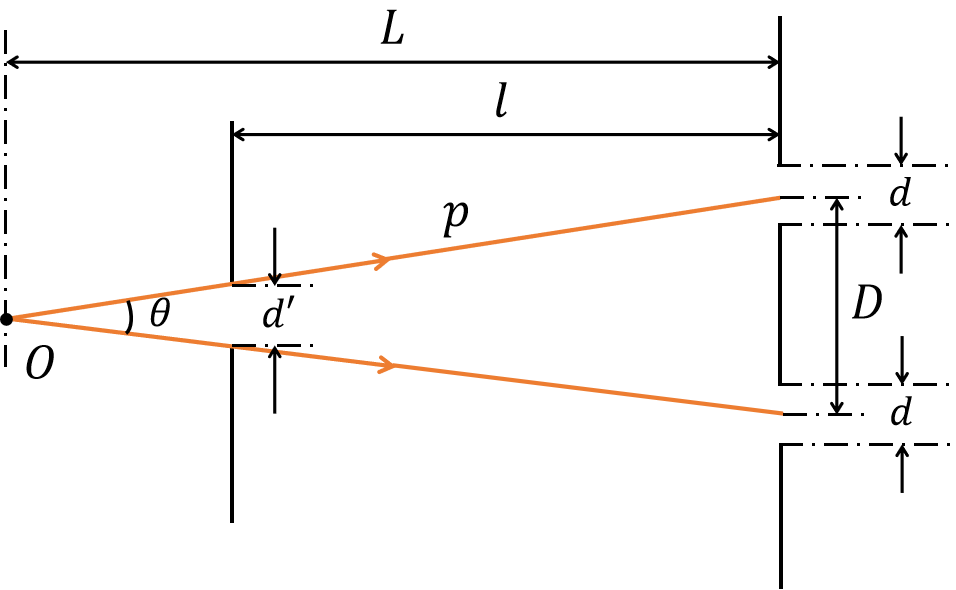}
	\caption{Parameters of the double-slit inteferometers.}
	\label{slit}
\end{figure}

When it comes to one of the most fascinating experimental phenomena in quantum mechanics, we most likely think of matter-wave interferometer.
By combining Eqs. \eqref{mw-T} and \eqref{mw-tau} through the definition in Eq. \eqref{def-beta}, the macroscopicity measure of matter-wave interferometers is given by
\begin{equation}
    \beta=\frac{p\theta d}{8\hbar}.
\end{equation}
The parameters $p$, $\theta$, and $d$ are depicted in FIG. \ref{slit}. Note that $\theta$ can not be directly determined from the separation of two slits $D$ and the flying distance $l$ between the collimating slit and the opaque plate. Instead, $\theta$ should be estimated indirectly through the collimating slit width by $\theta=D/L=(D-d')/l$.

Certainly, the matter-wave interferometers also incorporate the single-slit and multi-slit configurations. In the single-slit case, the diffraction pattern reveals that all particles passing through the slit participate in the interference process. Consequently, one can safely take $\theta=d/L$, where $L$ is the flying distance of particles from the source to the slit plate. In multi-slit cases, the situation is a little bit of subtlety, since one should figure out which (momentum of) particles are involved in the interference. According to Kirchhoff's diffraction theory \cite{1999Born}, for the number of slits $N>2$, the period of the interference fringes is given by $n/(ND)$ rather than $n/D$, where $n\in \mathbb{N}$. However, the interference peaks between two $n/D$ fringes are relatively weak, requiring extremely high coherence contrast to observe additional fringes \cite{2021triple-slit}. In matter-wave interference experiments, signals with a fringe period smaller than the slit period are rarely observed. That is, most coherence effects arise from interference of particles passing through adjacent slits, so one can still adopt $\theta=D/L$ in this type of experiment.

\begin{table}
	\centering
        \caption{Parameters and corresponding macroscopicity $\beta$ of the matter-wave interferometers in literatures. }
	\begin{tabular}{| c | c | c | c | c | c | c |}
    \hline
		Object & $M$(u) & $\lambda$(pm) &  $d$($\mu$m) & $\theta$($\mu$rad) & $\beta$ & Literature \\ 
		\hline\hline
		
		\multirow{2}{*}{Electron} & \multirow{2}{*}{5.48$\times 10^{-4}$} & 50 & 0.062 & 0.58 & $5.6\times 10^{-4}$ &2013\cite{2013electron} \\
            \cline{3-7}
		& & 30 & 0.05 & 0.63 & $8.2\times10^{-4}$ & 2018\cite{2018electron}\\
		
		\hline
		\multirow{3}{*}{Neutron} & \multirow{3}{*}{1} & $1901\pm 70$ & 96 & 15.2 & $0.60\pm0.02$ &  \multirow{2}{*}{1988\cite{1988neutron}} \\
        \cline{3-6}
		& & $1845\pm 280$ & 22.3 & 21.3  & $0.20\pm 0.03$ &  \\
        \cline{3-7}
		& & 500\footnote{The velocity range is inferred from the Maxwell-Boltzmann distribution in \cite{2018neutron}, which ultimately gives the range of $\beta$.} & 1.2 & 0.51 & $(4.7 - 14.3)\times 10^{-4}$ &2018\cite{2018neutron} \\
		
		\hline
		\multirow{2}{*}{Helium} & \multirow{2}{*}{4} & 
$33-100$ & 1 & 2.85 & $0.023 - 0.067$ & 1997\cite{1997helium} \\
        \cline{3-7}
		& & $30 - 60$ & 2 & 3.58 & $0.09 - 0.19$ & 1997\cite{1997helium2}\\
		
		\hline
		\multirow{7}{*}{Molecular} & 720 & $1.9 - 3.6$ & 0.05 & 0.088 & $0.0010 - 0.0018$ &1999\cite{1999c60} \\
        \cline{2-7}
		& 2814 & $1.1 - 2.3$ & \multirow{2}{*}{0.133} & \multirow{2}{*}{2.53} & $0.12 - 0.24$ & \multirow{2}{*}{2011\cite{2011mole}} \\
        \cline{2-3}\cline{6-6}
         & 5672 & $0.9 - 1.4$ & & & $ 0.19 - 0.29 $ &  \\
        \cline{2-7}
		& 515 & $2.5 - 4.4$ & 0.08 & 0.13 & $ 0.0018 - 0.0034$ & 2017\cite{2017mole} \\
        \cline{2-7}
		& 25000 & $0.061\pm 0.002$ & 0.133 & 0.27 & $0.46\pm 0.02$ & 2019\cite{2019mole} \\
        \cline{2-7}
		& 1882 & $0.354\pm 0.001$ & 0.0394 & 6.6 & $0.574\pm0.001$ & 2020\cite{2020mole} \\
        \cline{2-7}
		& 514.5 & $1.8 - 5.5$ & 0.08 & 0.13 & $0.0014 - 0.0044$ & 2021\cite{2021mole} \\
		\hline
	\end{tabular}
	\label{data}
\end{table}

With the de-Broglie relation $p=2\pi\hbar/\lambda$, one can obtain an equivalent description $\beta=\pi\theta d/4\lambda$. Table \ref{data} summarizes the experimental parameters and corresponding macroscopicity of various matter-wave interferometers reported in previous studies. The ranges of wavelength represent the possible velocity distributions of particles under thermal equilibrium. Within the coherence length of the source, a finite range of wavelength remains capable of inducing the interference patterns. To ensure rigorousness, we employ all possible momentum to determine the corresponding macroscopic parameter $\beta$. As for the slit parameters $d$ and $D$ (typically are some discrete sets in an individual experiment), we select the parameter sets yielding maximum $\beta$ values for analysis.

\begin{figure}
	\centering
    \subfigure
    {\includegraphics[width=0.65\linewidth]{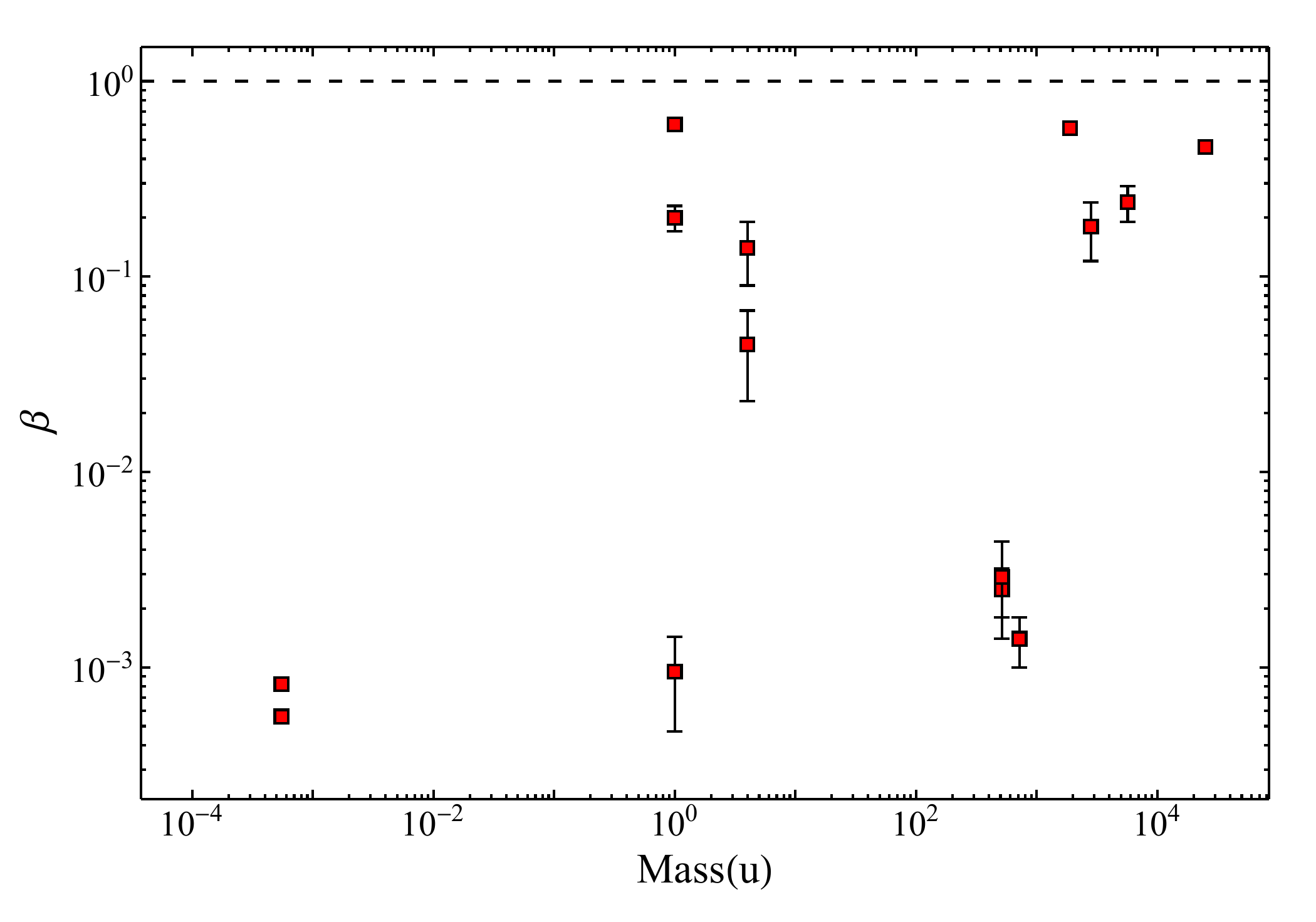}}
    \\
    \subfigure
    {\includegraphics[width=0.65\linewidth]{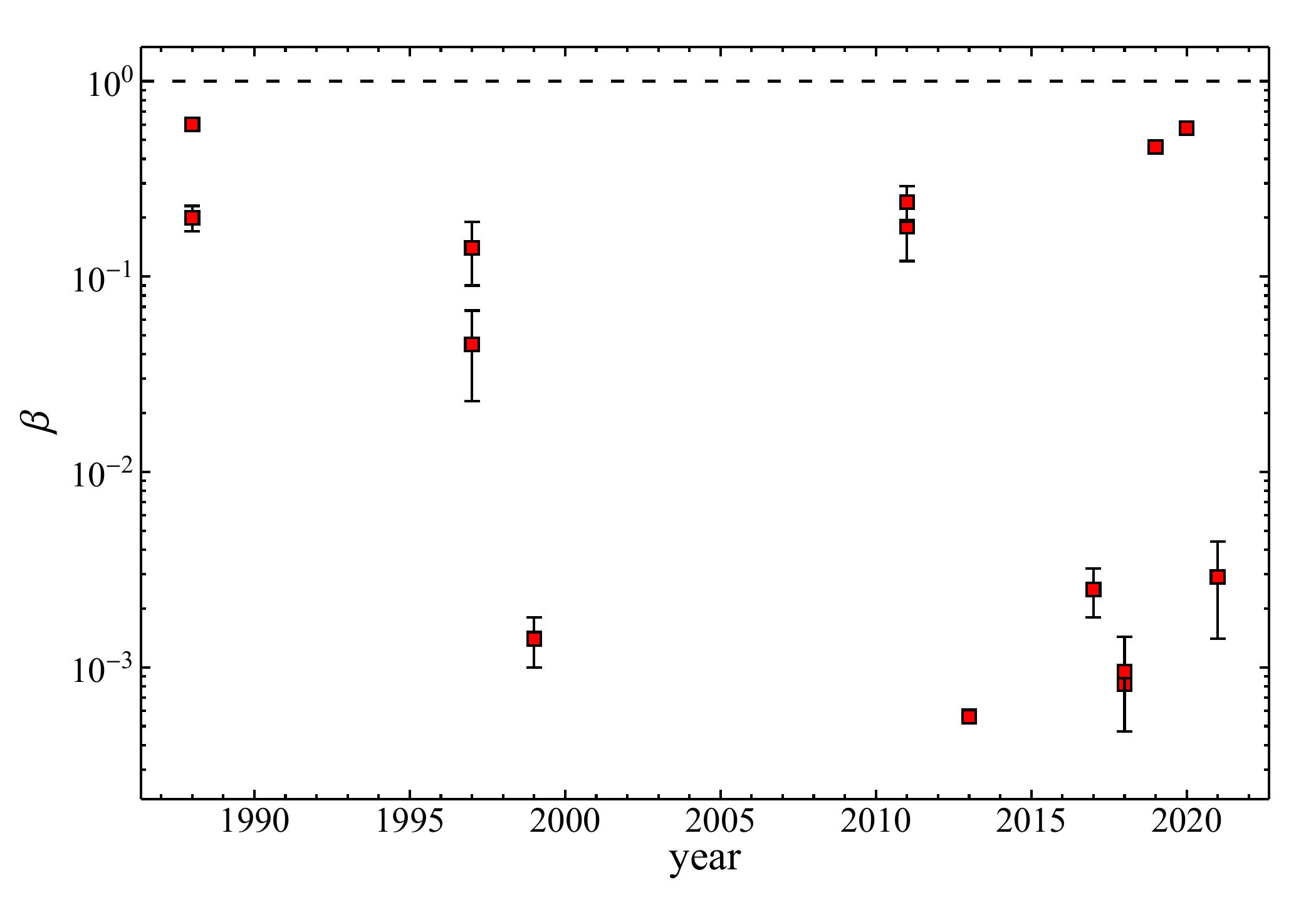}}        
	\caption{The macroscopicity measure $\beta$ varying with mass and year in the matter-wave interferometers. The plot consistently yields $\beta<1$, showing that, despite increasing mass scales, little advancement has been made in the exploration of quantum macroscopicity in recent years.}
	\label{fig}
\end{figure}

In order to emphasize that the Q-C boundary proposed in this paper differs from conventional intuition, the mass of each particle is also provided in Table \ref{data}, expressed in atomic mass units (u). The relationship between $\beta$ and $M$ is illustrated as a scatter plot in FIG. \ref{fig}. It can be observed that as the mass of the macromolecule increases, the corresponding macroscopicity remains consistent with $\beta<1$. Additionally, we also show how $\beta$ varies with the experiment year in FIG. \ref{fig}. It is not difficult to find out from the figure that, within the macroscopicity measure proposed in this paper, people have not made much progress in the exploration of Q-C boundary in recent years.

Now it is indispensable to evaluate the macroscopic extent of matter-wave interferometry in other indicators.
For instance, the disconnectivity $D$, introduced by Leggett \cite{1980Leggett}, has reached values as high as $D\sim N=2000$ with the development of modern quantum technology. The absence of a clear upper bound for $D$ makes it unsuitable as a definitive indicator for the Q–C boundary. The interference utility indicator, defined as $M=\theta_{\rm sing}/\theta_{\rm sup}$ in Ref. \cite{2004Bjork}, yields $M\simeq 5.2$ for the superposition of C$_{60}$ molecular in Ref. \cite{1999c60}, which looks good. However, this indicator only reflects macroscopicity after coherence has been observed and cannot predict superposition a priori. Another proposal is $\mu=\log_{10}(\tau_e/1{\rm s})$, based on minimal macrorealist modifications \cite{2020macro},  which gives a maximum value of $\mu\simeq 14$ for superpositions involving molecular masses above 25,000 u. However, the logarithmic scaling makes this measure increasingly difficult to test in practice.

Remarkably, we have deliberately excluded the momentum state coherence processes observed in the atomic fountain experiments and the neutron beam splitting interferometry \cite{1983Greeberger}. This exclusion is justified because both techniques employ fundamentally different configurations and parameter regimes compared to matter-wave interferometry. As a result, it becomes challenging to determine the momentum uncertainty of particles involved in the interference pattern.

\section{Macroscopicity of Ramsey-like Interferometer}\label{atomic fountain}

Ramsey's method is the basic way to determine a quantum superposition and its duration \cite{Ramsey1950}. The key point is similar to matter-wave interference: Two ``separate" beams of particles acquire distinct phases and then are recombined to observe variations in their population number, but this time it varies with frequency instead of position. Such a method applied in atomic fountains is of particular interest since the longer superposition duration generally implies more precise measurement in practice \cite{2001Peter}. 

By combining Eq. \eqref{Ram-tau} and experimentally observed coherence time $T$ through the definition in Eq. \eqref{def-beta}, we arrive at the macroscopicity measure for the Ramsey-like interferometers
\begin{equation}\label{empirical_indicator}
    \beta=\frac{8\pi^3\alpha(r_s\nu_s)^2}{9c^2}\nu_0T.
\end{equation}
The parameters $T$ and $\nu_0$ can be directly obtained from Ramsey spectroscopy: 1) Coherence time $T$ is the time interval between two Ramsey pulses; 2) $\nu_0$ is the center frequency in Ramsey fringe. For the system parameters $\nu_s$ and $r_s$, 3) we employ the transition frequency of the $D_1$ line as the system energy $\nu_s$. 4) The system size $r_s$ can be determined from $\nu_s$ through the transition probability of $D_1$ line $A_{fi}=32\alpha\pi^3\nu_s^3r_s^2/(9c^2)$.

The parameters of alkali metal atoms used in this study, which are widely adopted by the National Institute of Standards and Technology (NIST) in Ref. \cite{NIST}, are presented in Table \ref{Comparion}. The fifth column in Table \ref{Comparion} lists some typical superposition duration of corresponding atoms reported in recent literature. Then using these parameters and durations, we can determine the macroscopicity $\beta$ in the respective experiments. 

\begin{table}[!htbp]
	
	\begin{center}
		\caption{The macroscopicity measure $\beta$ for Ramsey-like interferometers in the literature.}
		\begin{tabular}{|c|c|c|c|c|c|c|}
			\hline
			\makebox[1.6cm][c]{Atom} & \makebox[1.6cm][c]{$\nu_0$/GHz}  & \makebox[1.6cm][c]{$\nu_s$/THz} & \makebox[1.6cm][c]{$r_s/a_0$} & \makebox[2.4cm][c]{$T$} & \makebox[1.6cm][c]{$\beta$} & \makebox[2.0cm][c]{Literature} \\
			\hline\hline
			
		    $^{23}$Na & 1.8 & 508 & 4.3 &  255ms & 14 & 1989 \cite{1989Kasevich} \\
			
			\hline
			
			\multirow{2}{*}{$^{39}$K} & \multirow{2}{*}{0.46} & \multirow{2}{*}{389} & \multirow{2}{*}{5.0} & 0.56ms & 0.006 & 1950 \cite{Ramsey1950} \\
			\cline{5-7}
			& & & & 40ms & 0.44 & 2017 \cite{2017K} \\
			
			\hline
			
			\multirow{2}{*}{$^{85}$Rb} & \multirow{2}{*}{3.0} & \multirow{2}{*}{377} & \multirow{2}{*}{5.2} & 142ms & 10 & 2015 \cite{2015_85Rb_87Rb} \\
			\cline{5-7}
			& & & & 1.8s & 130 & 2018 \cite{2018_85Rb_87Rb} \\
			
			\hline
			
			\multirow{6}{*}{$^{87}$Rb} & \multirow{6}{*}{6.8} & \multirow{6}{*}{377} & \multirow{6}{*}{5.2} & 0.5s & 81 & 1999 \cite{1999Rb} \\
			\cline{5-7}
			& & & & 2.3s & 370 & 2013 \cite{2013Rb} \\
			\cline{5-7}
			& & & & 600ms & 98 & 2013 \cite{2013Rb_Hu} \\
			\cline{5-7}
			& & & & 2.08s & 340 & 2015 \cite{2015Kovachy} \\
			\cline{5-7}
			& & & & 1.8s & 290 & 2018 \cite{2018_85Rb_87Rb} \\
			\cline{5-7}
			& & & & 280ms & 46 & 2023 \cite{2023Rb} \\
			
			\hline
			
			\multirow{3}{*}{$^{133}$Cs} & \multirow{3}{*}{9.2} & \multirow{3}{*}{335} & \multirow{3}{*}{5.5} & 0.354s & 70 & 1993 \cite{1993Cs} \\
			\cline{5-7}
			& & & & 0.6s & 119 & 2014 \cite{2014Cs} \\
			\cline{5-7}
			& & & & 801ms & 160 & 2018 \cite{2018Cs2} \\
			
			\hline

		\end{tabular}
	\label{Comparion}
	\end{center}
\end{table}

Notably, Ref. \cite{Ramsey1950} documents the original Ramsey experiment, which measured the hyperfine structure of $^{39}$K in 1950; Refs. \cite{1989Kasevich,1999Rb,1993Cs} were among the first to employ fountain experiments with alkali metal atoms for precision measurements. Therefore, these experimental data are listed additionally in the table \ref{Comparion}. In fountain experiment with both $^{85}$Rb and $^{87}$Rb atoms \cite{2018_85Rb_87Rb}, they exhibit the same coherence time. The primary physical motivation for this experiment was to test the weak equivalence principle between different isotopes. In this study, we find that the difference in the hyperfine energy gap $\nu_0$, caused by the nuclear structure of the two atoms, results in stronger quantum macroscopicity for $^{87}$Rb under the same coherence time. This difference suggests that future investigations into the Q-C boundary should focus more on $^{87}$Rb.

\begin{figure}
    \centering 
    \includegraphics[width=0.65\linewidth]{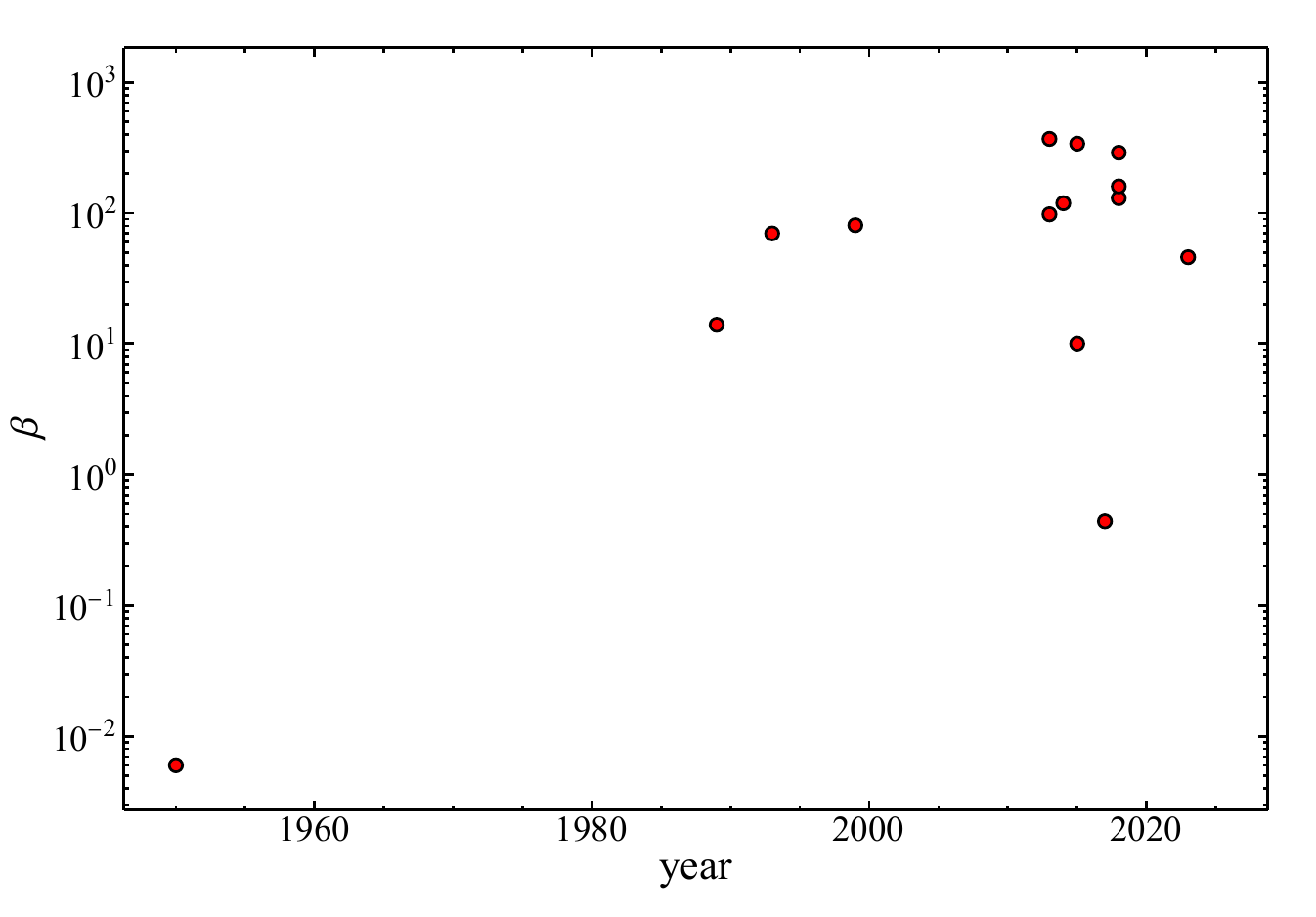}     
	\caption{The macroscopicity $\beta$ varying with the experimental year for Ramsey-like interferometers. The trend reflects a significant advancement in extending the Q-C boundary within the microscopic regime.}
	\label{fig2}
\end{figure}

We also present the variation of $\beta$ with the year of the experiments in FIG. \ref{fig2}. The increase of experimentally observed coherence time $T$, from the original Ramsey experiment to the modern cold atomic fountain experiment, directly leads to a corresponding increase in the macroscopicity parameter $\beta$. This trend reflects that we have made a significant advancement in extending the Q-C boundary within the microscopic regime.

It can be found that the largest macroscopicity, $\beta\sim 300$, is observed in state-of-the-art experiments reported in Refs. \cite{2015Kovachy,2018_85Rb_87Rb,2013Rb}, all of which were conducted using a vertical atomic fountain interferometer. One of the primary factors limiting the superposition duration, and consequently the macroscopicity $\beta$, is the practical height of the atomic fountain under Earth's gravity. To improve the coherence time, one can establish such interferometer in microgravity \cite{2013microgravity}, in space \cite{2021space}, or just construct taller interferometers \cite{2020ZAIGA}. Notably, if the macroscopicity $\beta \sim 300$ indeed represents the vicinity of the Q-C boundary, then any further attempts to extend coherence time will ultimately fail, regardless of how much effort we make to mitigate decoherence effects.

\section{Conclusion and Prospect}\label{conclusion}
This work provides a universal measure $\beta$ to quantify the macroscopicity in quantum superposition, through the ratio between the experimentally observed coherence time $T$ and the characteristic time $\tau$ required to repeatably distinguish the components of the superposition state, expressed as $\beta=T/\tau$. Unlike traditional approaches that rely on a single physical quantity, such as mass, separation of wave packets, or entangled particle numbers, our measure spans a feasible parameter space and thus provides a more unique characterization of the Q–C boundary.

For a specific matter-wave interferometer, the measure of its macroscopicity can be expressed as $\beta= p\theta d/(8\hbar)$. Analysis of experimental data, especially those involving macromolecular interference, consistently yields values of $\beta < 1$. This result suggests that the measure $\beta$ may capture some essential features underlying the quantum-to-classical transition in matter-wave interferometers. Therefore, we encourage experimentalists to revisit that failed coherence data, as some decoherence effects might not arise from environmental interactions, but instead from intrinsic limitations governed by fundamental principles.
Furthermore, we emphasize that this indicator can be obtained empirically from interferometry experiments. As such, it may remain valid even if the original model proposed in Ref. \cite{R2model} is eventually proven incorrect.

The analysis also shows that the measure of macroscopicity for atomic fountains can be expressed as $\beta=8\pi^3\alpha\nu_0r_s^2\nu_s^2T/ (9c^2)$. This challenges the conventional assumption that microscopic states are always adequately described by standard quantum mechanics. Remarkably, our analysis reveals that atomic fountain experiments can exhibit greater macroscopicity than macromolecular interferometers, i.e. $\beta\sim 300$ through concrete experiments data.

However, if the future experiments show that $\beta$ increases indefinitely with the development of experimental technology, just as the indicators in mass, system size, and degrees of freedom cases do, then a more universal measure may be required. This insight provides additional scientific motivation for developing precise atomic fountain experiments and offers a novel approach to exploring the Q-C boundary.

Additionally, for the superposition of particles localized in distinct potential traps, the measure of its macroscopicity is expressed as:
\begin{equation}
    \beta=\frac{ED}{4\pi\hbar c},
\end{equation}
where $E$ denotes the energy associated with the potential well and $D$ represents the spatial separation between the traps. Conceptually, this type of experiment can be interpreted as pixels in the measurement process. As such, testing this quantity may offer a novel and possibly final approach to addressing the quantum measurement problem.

\appendix
\section{Repeatably Distinguishing Time} \label{rdt}

In this appendix, we present, as simply as possible, how to repeatedly distinguish the components of a superposition in the matter-wave and Ramsey-like interferometers, and derive the characteristic time $\tau$ required in these processes.

\subsection{Matter-wave interferometer}
In the context of a matter-wave interferometer, the components of a superposition state correspond to two distinct trajectories of a free particle. We can distinguish them through the four-momentum conservation when scattering with photons (this scattering does not have to actually occur). To simplify the discussion, we adopt the approximation of total reflection, i.e., the scattered angle is $\pi$, which is just the Doppler-speed meter. 

Using the setup in Sec. \ref{interferometer}, in order to distinguish the two momenta associated with the two trajectories, we must have
\begin{equation}
\frac {Mc}{2} \cdot \frac {\Delta \omega }{\omega }  \le  \frac 12 p\cdot\theta .
\label{mw-distinction}
\end{equation}
The left-hand side of this inequality is the measurement error, which comes from the frequency uncertainty $\Delta\omega$ of the probe photon. Remarkably, according to the energy-time uncertainty relation, the minimum interaction duration is $\tau_{\rm min}\sim 1/\Delta\omega$. The right-hand side denotes the momentum difference of the two trajectories, namely the distinction.  

Apparently, the particle's momentum cannot be repeatedly distinguished in this way, since such a measurement does perturb the particle's momentum by an amount $2\hbar\omega/c$.
However, if this perturbation lies below the momentum uncertainty of those particles passing through the slits, then it is undetectable. Or one can say this perturbation can be neglected and distinguishment can be repeatedly operated. To this end, we require
\begin{equation}
\frac{2\hbar\omega}{c} \le p\frac{d}{2\sqrt{L^2+D^2/4}}\simeq p\frac{d}{2L} .
\label{mw-repeatable}
\end{equation}

Combining two inequalities \eqref{mw-distinction} and \eqref{mw-repeatable} to eliminate the dependence of photon frequency $\omega$, and noting that the full distinguishing process involves both emission and detection of photons—thus requiring twice the minimum interaction time $\tau_{\min}$—we arrive at the characteristic time given in Eq. \eqref{mw-tau}
\begin{equation}
    \tau=\frac{8\hslash L}{vp\theta d}.
\end{equation}

\subsection{Ramsey-like interferometer}

For the Ramsey-like interferometer, the typical components of a superposition are the hyperfine structure an atom. Here, we employ the total elastic scattering method to do the repeated distinguishment task, since there is no direct classical analogy as in the matter-wave interferometer case. 

Our starting point is the transition matrix describing the scattering of photons from hyperfine states $|F=0\rangle$ and $|F=1\rangle$ in the ground state of the Hydrogen atom (which can be straightforwardly generalized to hydrogen-like atoms). Without delving into the full derivation, the elastic scattering cross-section can be uniformly expressed as 
\begin{equation}\label{1S_cross_section}
	\sigma(F,\nu)\simeq \frac{16\pi^3\alpha^2}{3c^ 2}\cdot\nu^4\left|\sum_n \frac{|r_{n1}|^2}{3}\frac{-2\nu_{n1}+2F\cdot\nu_0}{(\nu_{n1}-F\cdot\nu_0)^2-\nu^2}\right|^2,
\end{equation}
where the argument $\nu$ denotes the frequency of scattering photons, $\nu_0$ is the hyperfine energy splitting, and $\nu_{n1}, r_{n1}$ are the transition frequency and radial matrix elements, respectively. The summation over $n$ represents the contribution of the possible transition with the intermediate states. Then the distinction between states can be defined through the difference in the cross-section
\begin{equation}\label{distinction}
	\mathcal{P}_d \propto \delta\sigma\equiv|\sigma(1,\nu)-\sigma(0,\nu)|.
\end{equation}

To establish the condition for repeatable distinguishment, we should first figure out the sources of inelastic processes. The most significant among these is the detuning absorption. Desipite we focus on scattering in the non-resonant regime, the finite duration of interaction induces the energy-time uncertainty, which still permits the absorption of detuned photons. The probability of such an absorption event is given by
\begin{equation}\label{Pabsorption}
    \mathcal{P}_a\propto\nu\sum_n \frac{|r_{n1}|^2}{(\nu_{n1}-\nu)^2}\sin^2[(\nu_{n1}-\nu)\tau/2].
\end{equation}

Note that the distinction probability $\mathcal{P}_d$ in \eqref{distinction} increases proportionally with the total duration $\tau$ of this scattering process, while the absorption probability $\mathcal{P}_a$ oscillates with $\tau$. Therefore, by imposing the condition $\mathcal{P}_d\geq \mathcal{P}_a$, which ensures that the distinction can be performed repeatably without destroying the state, we naturally obtain an expression for the characteristic time $\tau$, though it takes a relatively complex form:
\begin{equation}
    \tau=\frac{2\alpha\nu}{3\delta\sigma  }\sum_n\frac{|r_{n1}|^2}{(\nu_{n1}-\nu)^2}.
\end{equation}
To simplify this expression for practical use, we introduce several approximations: retain only the dominant intermediate state with $n=2$; replace $\nu$ with $\nu_{21}/2$; identify $|r_{21}|$, $\nu_{21}$ with the size $r_s$ and energy $\nu_s$ the system, respectively. It is worth noting that any overestimation introduced by these approximations can be absorbed into the macroscopicity parameter $\beta$.  With these simplifications, we arrive at the repeatable distinguishing time given in Eq. \eqref{Ram-tau}.

\bibliographystyle{apsrev4-1}
\bibliography{indicator_reference}

\end{document}